\documentclass[conference]{IEEEtran}
\IEEEoverridecommandlockouts
\usepackage{cite}
\usepackage{amsmath,amssymb,amsfonts}
\usepackage{algorithmic}
\usepackage{graphicx}
\usepackage{textcomp}
\usepackage{comment}
\usepackage{xcolor}
\usepackage{url}
\usepackage{hyperref}
\usepackage{booktabs}

\def\BibTeX{{\rm B\kern-.05em{\sc i\kern-.025em b}\kern-.08em
    T\kern-.1667em\lower.7ex\hbox{E}\kern-.125emX}}
\begin{document}

\title{Inverse Modeling of Dielectric Response in Time Domain using Physics-Informed Neural Networks\\
\vspace{-0.3em}
}

\author{
\small
E. Esenov\textsuperscript{1*}, O. Hjortstam\textsuperscript{1,2}, Y. Serdyuk \textsuperscript{1}, T. Hammarström\textsuperscript{1}, C. Häger\textsuperscript{1} \vspace{0.15em} \\
\textsuperscript{1}Chalmers University of Technology, Gothenburg, Sweden \\
\textsuperscript{2}Hitachi Energy Research, Västerås, Sweden \vspace{0.15em} \\
*emir.esenov@chalmers.se
}
\maketitle

\begin{abstract}
Dielectric response (DR) of insulating materials is key input information for designing electrical insulation systems and defining safe operating conditions of various HV devices. In dielectric materials, different polarization and conduction processes occur at different time scales, making it challenging to physically interpret raw measured data. To analyze DR measurement results, equivalent circuit models (ECMs) are commonly used, reducing the complexity of the physical system to a number of circuit elements that capture the dominant response. This paper examines the use of physics-informed neural networks (PINNs) for inverse modeling of DR in time domain using parallel RC circuits. To assess their performance, we test PINNs on synthetic data generated from analytical solutions of corresponding ECMs, incorporating Gaussian noise to simulate measurement errors. Our results show that PINNs are highly effective at solving well-conditioned inverse problems, accurately estimating up to five unknown RC parameters with minimal requirements on neural network size, training duration, and hyperparameter tuning. Furthermore, we extend the ECMs to incorporate temperature dependence and demonstrate that PINNs can accurately recover embedded, nonlinear temperature functions from noisy DR data sampled at different temperatures. This case study in modeling DR in time domain presents a solution with wide-ranging potential applications in disciplines relying on ECMs, utilizing the latest technology in machine learning for scientific computation.
\end{abstract}


\section{Introduction}
Dielectric materials are central to modern electrical and electronic systems, from power transmission and high-voltage insulation to capacitor film materials. Accurately characterizing their dielectric response (DR) is necessary for assessing material performance, reliability, and degradation under operational stresses. However, interpreting DR data to extract meaningful physical parameters remains a challenging inverse problem. Specifically, methods for inverse modeling of physical parameters--such as conductivity, relaxation times, and activation energies--with nonlinear temperature field dependence and noisy data are inherently complex, underscoring the demand for improved modeling techniques \cite{zaengl, dai}.

PINNs are an emerging computational framework integrating governing physical laws into neural network (NN) training. Unlike purely data-driven models, PINNs incorporate domain knowledge such as differential equations and initial/boundary conditions to guide the learning process, reducing the dependency on large labeled datasets and improving physical consistency \cite{pinn}. This approach is also helpful for solving inverse problems in physics, where measured data is often sparse and noisy, yet underlying physical principles impose strong constraints on valid solutions. 

In a recent study~\cite{dai}, conventional fitting methods were applied to extract equivalent circuit parameters from DR measurements of various insulation materials, including those operating above and below their glass transition temperatures. That work suggested that future use of AI techniques, particularly PINNs, could enhance the accuracy and efficiency of DR analysis. More recently, PINNs have been applied to model ionic charge transport in high-voltage gas insulation~\cite{arni, carl-j}, demonstrating the potential to overcome known numerical limitations of the finite element method and to efficiently extract physical parameters from measured data. However, using PINNs to analyze DR data has not yet been reported in the literature to the authors' knowledge.

This paper proposes an inverse modeling approach for DR analysis in time domain using PINNs. Our study focuses on the parallel RC equivalent circuit representation of dielectric materials and examines the efficacy of PINNs for estimating RC parameters from synthetic DR data. In addition to scalar parameter estimation, we conduct simulations extending the equivalent circuit to include nonlinear temperature effects. In this setting, we explore the ability of PINNs to recover temperature-dependent activation functions from DR data sampled at varying temperatures. By combining deep learning with dielectric physics, our method aims toward advancing computational techniques for analyzing next-generation dielectric materials with tailored properties.

\begin{figure*}[!t]
  \centering
  \includegraphics[width=0.9\textwidth]{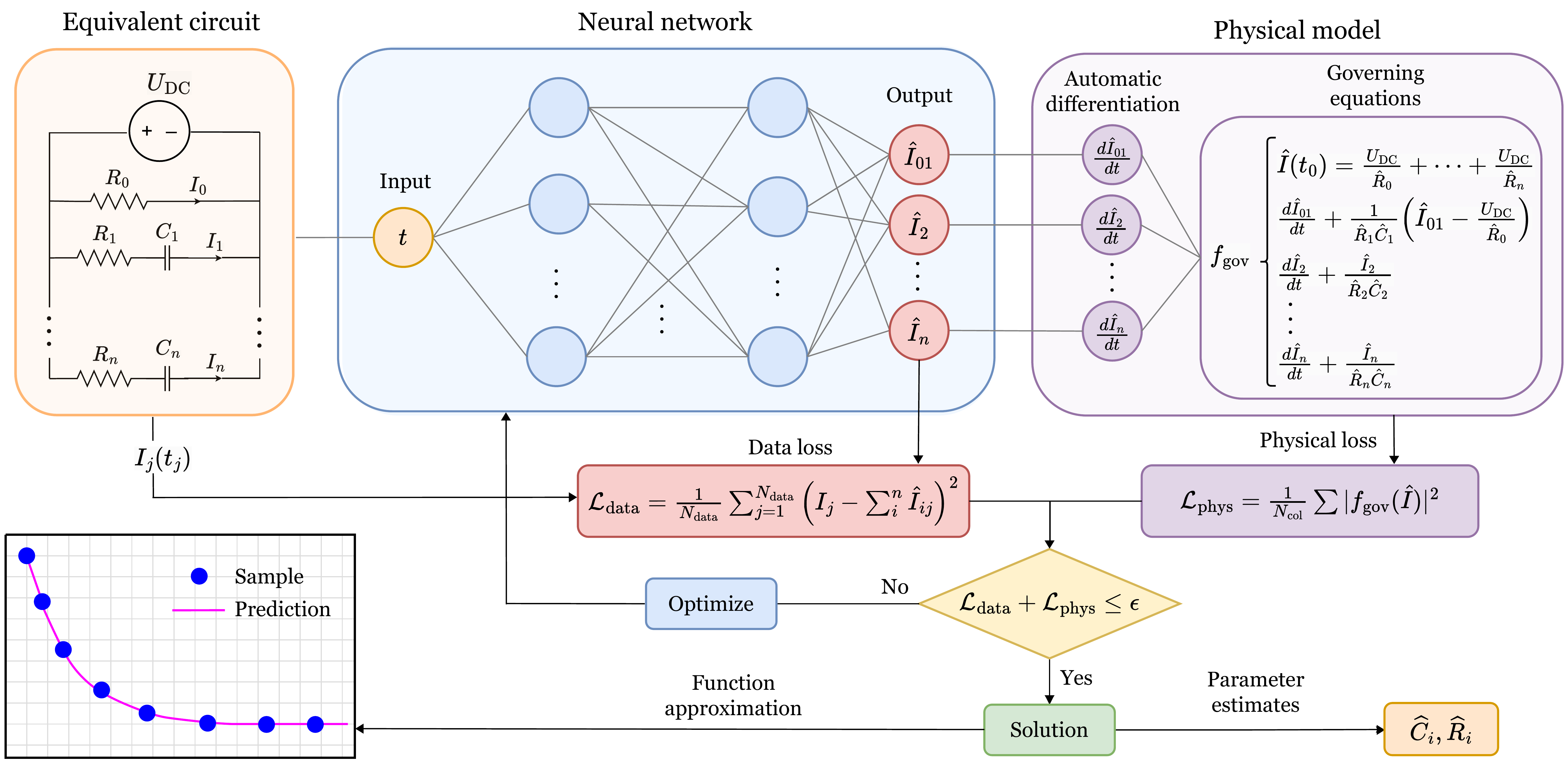}
  \caption{A schematic representation of inverse modeling of DR in time domain using PINNs.}
  \label{fig:project_figure}
\end{figure*}

\section{Methodology} \label{sec:methodology}

\subsection{Equivalent circuit model}

The DR of materials in time domain can be modeled using an equivalent circuit composed of parallel-connected RC elements as shown in Fig.~\ref{fig:project_figure}, each representing a dielectric relaxation process with its characteristic time constant\cite{kuchler2017high}. Resistance $R_0$ accounts for a steady-state conduction path. The circuit is stressed by applying a step voltage $U_{\text{DC}}$. The total current $I(t)$ is composed of both steady-state and transient components according to
\begin{equation} \label{eq:I_tot}
I(t) = \frac{U_{\text{DC}}}{R_0} + \sum_{i=1}^{n} \frac{U_{\text{DC}}}{R_i} \cdot \exp\left(-\frac{t}{R_i C_i}\right),
\end{equation}
where each exponential term represents the transient current through a corresponding RC branch. The system behavior can be equivalently described using a set of coupled first-order differential equations, where the dynamics of the different branches are given by:
\begin{equation} \label{eq:ODEs}
    \begin{aligned}
        \frac{dI_{01}}{dt} + \frac{1}{R_{1}C_{1}}\left( I_{01}-\frac{U_{\text{DC}}}{R_{0}} \right) = 0, \\
    \frac{dI_{i}}{dt} + \frac{I_{i}}{R_{i}C_{i}} = 0, \quad \text{for} \: i=2,\dots,n.
    \end{aligned}
\end{equation}

\begin{equation} \label{eq:IC}
I(t=0) = \frac{U_{\text{DC}}}{R_0} + \sum_{i=1}^{n} \frac{U_{\text{DC}}}{R_i}
\end{equation}
with an initial condition assuming a system with no residual polarization or pre-existing capacitance charging.

\subsubsection{Temperature effects}

To incorporate temperature effects, we extend the governing system~(\ref{eq:I_tot})--(\ref{eq:IC}) by modeling each resistive component $R_i$ as a function of temperature. Following the approach in~\cite{Ho2012HighFC}, we assume an Arrhenius-type expression
\begin{equation} \label{eq:temp}
    R_i(T) = A_i \cdot e^{\frac{W}{k_BT}},
\end{equation}
where $W=0.76$ eV is the activation energy, $k_B$ is the Boltzmann constant, and $A_i$ is a scaling constant specific to each resistive element. These values are selected based on the referenced model and reflect the typical behavior of temperature-dependent dielectric materials.

\subsection{PINN model}

PINNs~\cite{pinn} are an extension of conventional NNs~\cite{Schmidhuber_2015} that augment the training process by directly incorporating domain knowledge, such as differential equations and initial/boundary conditions, into the loss function. Whereas conventional NNs usually rely solely on data, the PINN extension allows the network to use the known structure of the underlying physical problem during training. 

The application of PINNs in this work for inverse modeling of DR without temperature-dependence is illustrated in Fig.~\ref{fig:project_figure}. The training data is sampled from the analytical solution of the equivalent circuit~\eqref{eq:I_tot} and consists of input-output pairs $(t,I)$ representing the total current measurement $I$ for a given time $t$. Time points are uniformly sampled over the interval $t \in [0, t_{\text{end}}]$, where $t_{\text{end}}$ is chosen such that the transient response has sufficiently decayed and the system has reached steady-state. To emulate measurement noise typically present in experimental data, Gaussian noise is added to the current samples such that $I^{\text{data}} = I + \mathcal{N}(0,\sigma^2)$, where $\mathcal{N}(0,\sigma^2)$ denotes zero-mean Gaussian noise with variance $\sigma^2$. The NN is trained to learn the functional relationship between time $t$ and the surrogate branch currents $\hat{I}_{01},\hat{I}_2,...,\hat{I}_n$ of the circuit. These outputs are constrained by the governing equations, enabling the network to infer polarization dynamics from the total current data. The training loss combines the data loss computed over all available training samples with the physical loss comprising governing equations evaluated at uniformly sampled collocation points in the time domain. Additionally, all surrogate circuit parameters $\hat{R}_i$ and $\hat{C}_i$ are jointly optimized along with the NNs' weights and biases during training.

In cases where temperature effects are considered, we simulate repeated current measurements at different temperatures, each held constant during its respective measurement. The training data is sampled as before, using resistances computed from~\eqref{eq:temp}. Each data point is represented as $(t,T,I),$ where $t$ is the time, $T$ is the temperature, and $I$ is the total current. To account for noise amplification commonly observed in low-current measurements, zero-mean Gaussian noise is added as $I^{\text{data}} = I + \mathcal{N}(0,\left(\sigma/I\right)^2)$. The objective is to estimate the circuit parameters $\hat{C}_i$ and $\hat{R}_i(T)$ from the data, where the temperature-dependent resistances $\hat{R}_i(T)$ are modeled using subnetworks. Specifically, the temperature input $T$ is first passed through a dedicated subnetwork that outputs $\hat{R}_i(T),$ which is then combined with the time input $t$ to the main network responsible for predicting the different branch currents. In cases involving multiple temperature-dependent resistances, a single subnetwork is used to learn the functional form of a representative resistance, while the remaining resistances are modeled as scaled versions of this function, with each scaling factor learned during training.

For DR inverse modeling to be well-conditioned, each RC branch must contribute a distinguishable signal to the total current, as individual branch currents are not part of the training data. This requires sufficient curvature and temporal separation between polarization components. When polarization responses are weak or overlapping, the total current becomes ill-conditioned, allowing multiple high-accuracy solutions. This is a generic limitation of the ECMs used in this work and not specific to PINNs. This work considers only well-conditioned DRs, where distinct polarization components enable reliable inverse modeling. 

Following standard machine learning practice, we operate with normalized input features and model parameters in appropriate numerical ranges, ensuring numerical stability and convergence during training. All results are reported in arbitrary units (a.u.) that do not directly correspond to real-world magnitudes. In practical applications, this normalization can be achieved by scaling physical values before training and rescaling afterward to recover the original dimensional quantities.

All models were implemented in Python using the JAX-PI library~\cite{wang2023expert}. We extended its functionality to support inverse problems in our modified version \cite{forkjaxpi}, along with subnet architectures for temperature-dependent simulations. Stable convergence did not depend on significant hyperparameter tuning, so we primarily used the library’s default settings: the Adam optimizer with an exponential decay learning rate schedule and tanh activation functions. Due to the small size of the training datasets, data batching was not applied. Collocation batch size was kept at the default value of 4096, which could be significantly reduced given the small time domain considered in this work; it was retained as it has negligible impact on runtime. Learning rates of either $1 \times10^{-2}$ or $1 \times10^{-1}$ were used depending on the numerical range of the estimated parameters. We recommend normalizing parameters to small numerical ranges combined with a lower learning rate for improved stability. Network architectures were kept minimal, consisting of one or two hidden layers with fewer than 30 neurons each. All models were trained in under 60 seconds on a standard-issue MacBook Air M3, highlighting the method’s low computational requirements. Code for all simulations is available at~\cite{github}.

\section{Results and Discussion} \label{sec:results}

\subsection{Static Parameter Recovery}

We first evaluate the PINN model's performance without temperature dependence. Among the tested configurations, we present results for the most complex case: an equivalent circuit consisting of one steady-state resistive branch and two polarization branches, yielding five unknown parameters.

The PINN architecture used in this task comprised two hidden layers with 15 neurons each. The network outputs correspond to the surrogate branch currents $\hat{I}_{01}$ and $\hat{I}_2$. The model was trained for 50,000 iterations on 50 synthetic data samples generated using ~\eqref{eq:I_tot}. To simulate experimental noise, we added Gaussian noise at varying levels ($\sigma = 0.0$, $0.1$, $0.2$) to the current signal.

\begin{figure}[t!]
    \centering
    \includegraphics[width=0.4\textwidth]{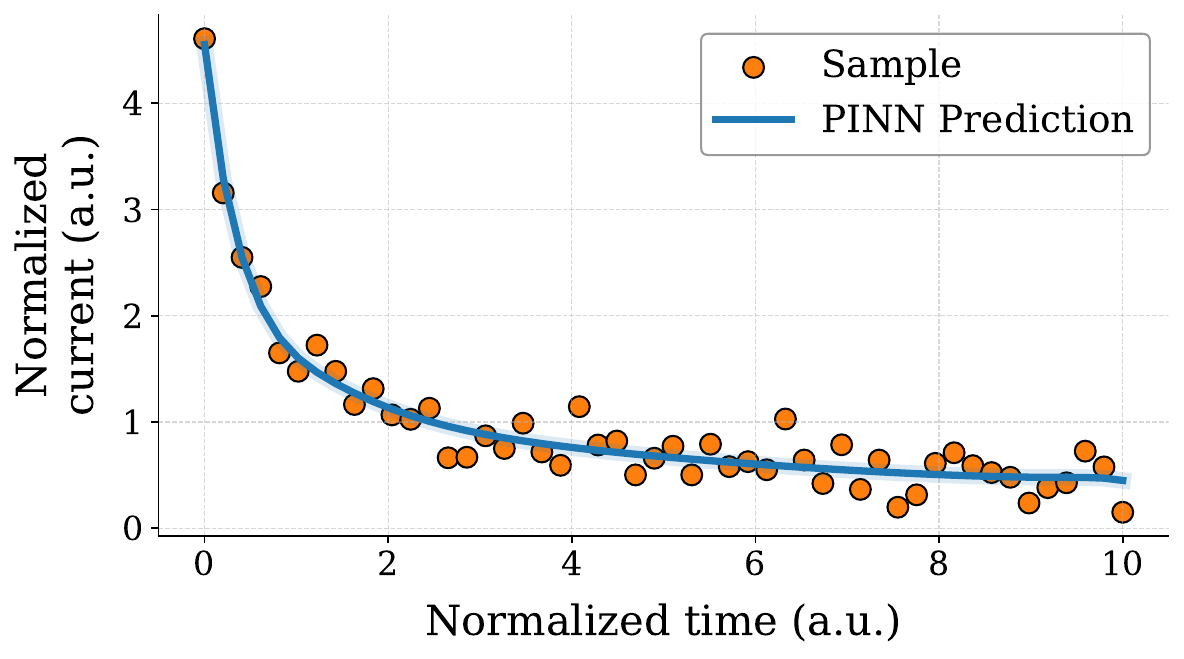}
    \caption{PINN-predicted dielectric response for an ECM comprising one steady-state branch and two RC branches, trained on synthetic data with added Gaussian noise \( \mathcal{N}(0,0.2^2) \).}
    \label{fig:case2sol}
\end{figure}

\begin{figure}[t!]
    \centering
    \includegraphics[width=0.48\textwidth]{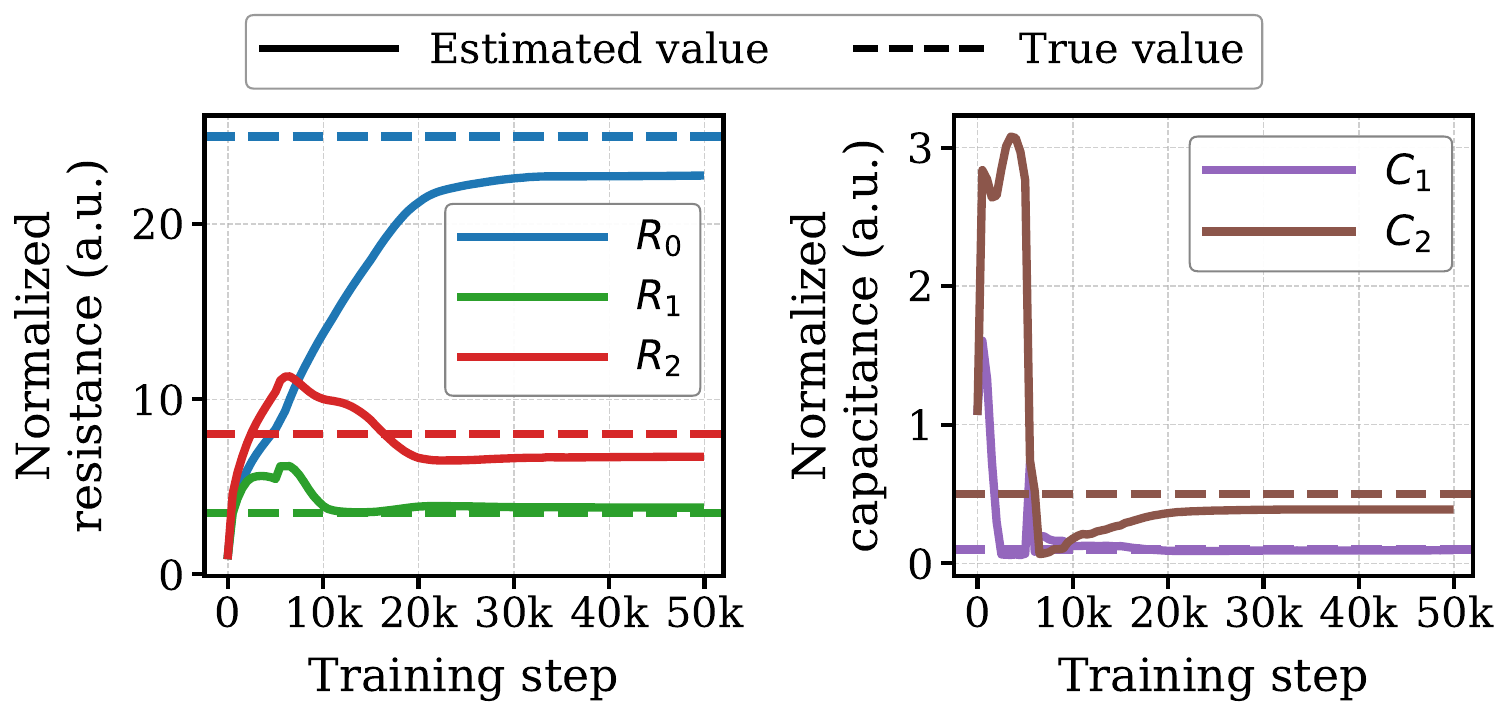}
    \caption{Convergence of estimated circuit parameters during training for five-parameter ECM with added Gaussian noise $\mathcal{N}(0,0.2^2)$.}
    \label{fig:case2param}
\end{figure}

Table~\ref{tab:case2} summarizes the recovered circuit parameters under different noise conditions. The results demonstrate that the PINN framework accurately estimates all five parameters, with modest degradation in performance as noise increases. The relative error remains within acceptable bounds even at a $\sigma = 0.2$ noise level. Fig.~\ref{fig:case2sol} illustrates the predicted DR curve under the highest noise level, showing strong agreement with the noisy ground truth. The convergence behavior of parameter estimates, at $\sigma = 0.2$, during training is shown in Fig.~\ref{fig:case2param}.

\begin{table}[b!]
\centering
\caption{Estimated circuit parameters under varying noise levels for one steady-state branch and two polarization branches.}
\label{tab:case2}
\begin{tabular}{lccccc}
\toprule
\textbf{Parameters} & \textbf{$R_0$} & \textbf{$C_1$} & \textbf{$R_1$} & \textbf{$C_2$} & \textbf{$R_2$} \\
\midrule
\textbf{True values} & 25.00 & 0.10 & 3.50 & 0.50 & 8.00 \\
\textbf{$\sigma = 0.0$} & 24.73 & 0.10 & 3.51 & 0.49 & 8.00 \\
\textbf{$\sigma = 0.1$} & 23.76 & 0.10 & 3.60 & 0.44 & 7.53 \\
\textbf{$\sigma = 0.2$} & 22.77 & 0.09 & 3.81 & 0.39 & 6.72 \\
\bottomrule
\end{tabular}
\end{table}

\subsection{Modeling Temperature Dependence}

\begin{figure}[b!]
    \centering
    \includegraphics[width=0.425\textwidth]{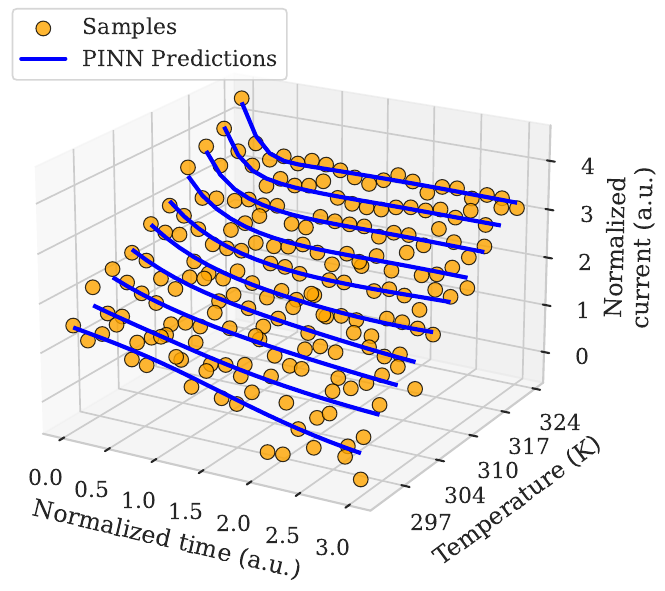}
    \caption{PINN-predicted DR across varying temperatures (294–324 K) for an ECM with one steady-state and one polarization branch. Results shown correspond to synthetic data with added Gaussian noise $\mathcal{N}(0,(0.4/I)^2)$.}
    \label{fig:case1sol}
\end{figure}

Next, we evaluate the performance of PINNs for estimating temperature-dependent circuit parameters. The test configuration consists of an ECM with one resistive branch and one polarization branch, leading to three unknowns: $\hat{C}_1$, $\hat{R}_0(T)$, and $\hat{R}_1(T)$. Here, $\hat{R}_1(T)$ is modeled as a scalar multiple of $\hat{R}_0(T)$, which is learned through a dedicated subnetwork.

Synthetic training data were generated for 10 uniformly spaced temperatures in the range 294–324 K. For each temperature, 20 current samples were produced, and Gaussian noise was added at three levels: $(\sigma = 0.0, 0.2, 0.4),$ where the current magnitude scales $\sigma$; see Section~\ref{sec:methodology}. The main and subnetwork architectures each used a single layer with 15 neurons, and training was performed for 50,000 iterations.

\begin{figure}[t!]
    \centering
    \includegraphics[width=0.4\textwidth]{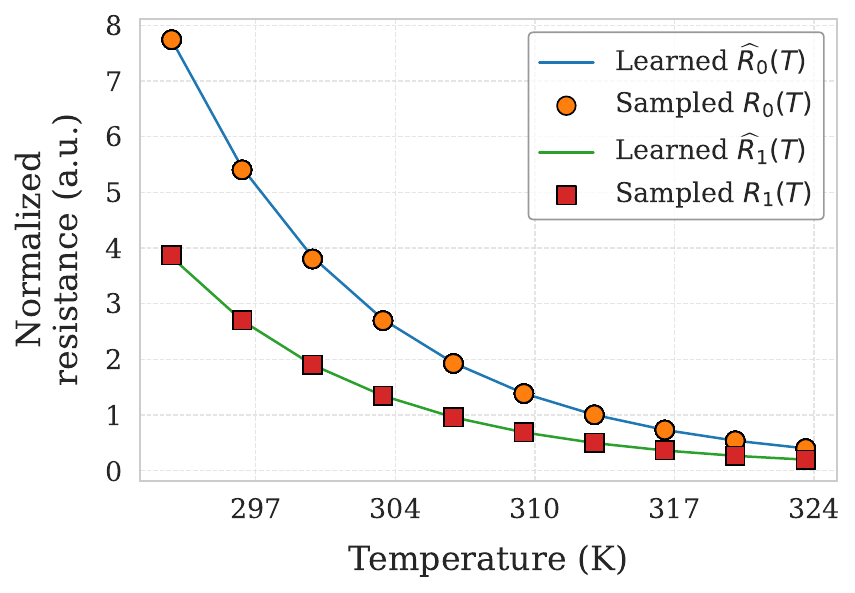}
    \caption{Learned functional approximations for \(\hat{R}_0(T)\) and scaled \(\hat{R}_1(T)\) with added Gaussian noise $\mathcal{N}(0,(0.4/I)^2).$}
    \label{fig:case1params}
\end{figure}

Fig.~\ref{fig:case1sol} presents the predicted DR curves across temperatures under the highest noise level ($\sigma = 0.4$), showing good agreement with the synthetic data. Table~\ref{tab:case1} reports the estimated scalar parameters across all noise levels. The recovered $\hat{R}_1$ scaling factor remains stable across increasing noise, while the capacitance estimate $\hat{C}_1$ shows some degradation. Fig.~\ref{fig:case1params} shows the learned temperature-dependent resistance functions, highlighting the capacity of PINNs to accurately model embedded nonlinear thermally activated resistances from the noisy DR data.

\begin{table}[h]
\centering
\caption{Estimated circuit parameters under varying noise levels for ECM with temperature-dependent resistance.}
\label{tab:case1}
\begin{tabular}{lcc}
\toprule
\textbf{Parameters} & \textbf{$R_1$ (scaling)} & \textbf{$C_1$} \\
\midrule
\textbf{True values}      & 0.50   & 0.50   \\
\textbf{$\sigma = 0.0$}   & 0.4986 & 0.5399 \\
\textbf{$\sigma = 0.2$}   & 0.4978 & 0.5445 \\
\textbf{$\sigma = 0.4$}   & 0.4956 & 0.5753 \\
\bottomrule
\end{tabular}
\end{table}

\section{Conclusion}

This work presents a PINN approach for inverse modeling of DR in time domain using equivalent RC circuits. We demonstrate that PINNs can accurately estimate circuit parameters from noisy synthetic data with minimal model complexity and training overhead. Furthermore, by incorporating temperature-dependent behavior into the equivalent circuit, our method successfully recovers nonlinear resistances from data measured at varying temperatures. These results highlight the potential of PINNs as a powerful tool for analyzing DR, offering a robust and computationally efficient alternative to conventional parameter extraction techniques. Future work will extend this framework to frequency domain, alternative ECM representations, and experimental datasets.

\bibliographystyle{ieeetr}
\bibliography{refs}

\end{document}